\documentstyle[12pt,psfig]{article}
\def\be{\begin{equation}}
\def\ee{\end{equation}}
\def\bea{\begin{eqnarray}}
\def\eea{\end{eqnarray}}

\begin{document}
\title{PARTICLE PRODUCTION FROM SIS TO SPS ENERGIES}
\author{ W. CASSING \\
\em Institut f\"ur Theoretische Physik,
\em Heinrich-Buff-Ring 16,\\
\em  D-35392 Giessen, Germany\\
\em E-mail: cassing@theorie.physik.uni-giessen.de}
\date{}

\maketitle

\begin{abstract}
The production and propagation of mesons ($\pi, \eta, \rho, \omega,
\Phi, K, \bar{K}, J/\Psi$) in proton-nucleus and nucleus-nucleus
collisions from 1 - 200 GeV/u is studied within the covariant transport
approach  HSD, which explicitly allows to investigate selfenergy
effects of the hadrons at finite baryon density. Whereas the
experimental pion and $K^+$ spectra can be described without
introducing any selfenergies for the mesons, the $K^-$ yield in Ni + Ni
collisions is underestimated by a factor of 5--7 at 1.66 and 1.85
GeV/u. However, introducing density dependent antikaon masses in line
with effective chiral Lagrangians a satisfactory agreement with the
data is achieved.  A dropping of the $\rho$-meson mass with baryon
density, as suggested by QCD sumrule studies, is proposed to explain
the dilepton spectra for S + Au and Pb + Au at SPS energies, which
indicates independently that a partial restoration of chiral symmetry
might be found already in the present experiments.
\end{abstract}

\section{Introduction}
The study of hot and dense nuclear matter via relativistic
nucleus-nucleus collisions is the major aim of high energy heavy-ion
physics. Nowadays, the search for a restoration of chiral symmetry at
high baryon density or for a phase transition to the quark-gluon plasma
(QGP) is of specific interest. However, any conclusions about the
hadron properties at high temperature or baryon densities must rely on
the comparison of experimental data with theoretical approaches based
on nonequilibrium kinetic theory. Among these, the covariant RBUU
approach~\cite{Cass,Weber1}, the QMD~\cite{Aich} or RQMD
model~\cite{RQMD}, and the HSD approach~\cite{Ehehalt} have been
successfully used in the past. As a genuine feature of transport
theories there are two essential ingredients: i.e.  the  baryon (and
meson) scalar and vector selfenergies -- which are neglected in many
approaches -- as well as in-medium elastic and inelastic cross sections
for all hadrons involved.

Selfenergy effects in the production of particles have been found so
far for antiprotons at SIS energies \cite{Teis,Ko3}, though the actual
magnitude of the attractive $\bar{p}$-potential in the nuclear medium
is still a matter of debate since the strong $\bar{p}$ annihilation
with nucleons is not sufficiently under control. As advocated in
Refs.~\cite{GB1,Brown,Kaplan,waas} on the basis of effective chiral
Lagrangians also antikaons should feel strong attractive forces in the
medium so that their production threshold should be reduced at finite
baryon density. The vector mesons $\rho$ and $\omega$, furthermore, are
expected to drop in mass with baryon density according to QCD sumrule
studies \cite{hatsuda}, which is basically a consequence of the
dropping scalar quark condensate $<\bar{q}q>$ with the quark density
$<q^{\dagger}q>$ \cite{hhh}. Whereas a direct enhancement of the
$\rho$-meson yield is hard to observe experimentally due to the short
lifetime of the $\rho$ and the strong final state interactions of the
pions, dilepton spectroscopy is expected to provide valuable
information on the $\rho$ spectral function in the dense medium.

The present article is organized as follows: Section 2 contains a brief
description of the transport approach employed as well as a
specification of the meson selfenergies incorporated in the
calculation.  Section 3 is devoted to a presentation of the calculated
results for $\pi, K^+,$ and $K^-$ spectra in comparison to available
data at SIS, AGS and SPS energies. Section 4 concentrates on dilepton
physics at SPS energies while Section 5 concludes this study with a
summary and discussion of open problems.

\section{Ingredients of the transport approach}

In this work the dynamical analysis of p+A and A+A reactions is
performed within the HSD \footnote{Hadron String Dynamics}
approach~\cite{Ehehalt} which is based on a coupled set of covariant
transport equations for the phase-space distributions $f_{h} (x,p)$ of
hadron $h$~\cite{Weber1,Ehehalt}, i.e.
\begin{eqnarray}  \label{g24}
\lefteqn{\left\{ \left( \Pi_{\mu}-\Pi_{\nu}\partial_{\mu}^p U_{h}^{\nu}
-M_{h}^*\partial^p_{\mu} U_{h}^{S} \right)\partial_x^{\mu}
+ \left( \Pi_{\nu} \partial^x_{\mu} U^{\nu}_{h}+
M^*_{h} \partial^x_{\mu}U^{S}_{h}\right) \partial^{\mu}_p
\right\} f_{h}(x,p) } \nonumber \\
&& = \sum_{h_2 h_3 h_4\ldots} \int d2 d3 d4 \ldots
 [G^{\dagger}G]_{12\to 34\ldots}
\delta^4(\Pi +\Pi_2-\Pi_3-\Pi_4 \ldots )  \nonumber\\
&& \times \left\{ f_{h_3}(x,p_3)f_{h_4}(x,p_4)\bar{f}_{h}(x,p)
\bar{f}_{h_2}(x,p_2)\right.  \nonumber\\
&& -\left. f_{h}(x,p)f_{h_2}(x,p_2)\bar{f}_{h_3}(x,p_3)
\bar{f}_{h_4}(x,p_4) \right\} \ldots\ \ .
\end{eqnarray}
In Eq.~(\ref{g24}) $U_{h}^{S}(x,p)$ and $U_{h}^{\mu}(x,p)$ denote the
real part of the scalar and vector hadron selfenergies, respectively,
while $[G^+G]_{12\to 34\ldots} \delta^4 (\Pi +\Pi_2-\Pi_3-\Pi_4\ldots
)$ is the 'transition rate' for the process $1+2\to 3+4+\ldots$ which
is taken to be on-shell in the semiclassical limit adopted. The hadron
quasi-particle properties in (\ref{g24}) are defined via the mass-shell
constraint~\cite{Weber1},
\begin{equation}   \label{g25}
\delta (\Pi_{\mu}\Pi^{\mu}-M_{h}^{*2} ) \ \ ,
\end{equation}
with effective masses and momenta (in local Thomas-Fermi approximation)
given by
\begin{eqnarray}\label{g26}
M_{h}^* (x,p)&=&M_h + U_h^{{S}}(x,p) \nonumber \\
\Pi^{\mu} (x,p)&=&p^{\mu}-U^{\mu}_h (x,p)\ \ ,
\end{eqnarray}
while the phase-space factors
\begin{equation}
\bar{f}_{h} (x,p)=1 \pm f_{{h}} (x,p)
\end{equation}
are responsible for fermion Pauli-blocking or Bose enhancement,
respectively, depending on the type of hadron in the final/initial
channel. The dots in Eq.~(\ref{g24}) stand for further contributions to
the collision term with more than two hadrons in the final/initial
channels. The transport approach (\ref{g24}) is fully specified by
$U_{h}^{S}(x,p)$ and $U_{h}^{\mu}(x,p)$ $(\mu =0,1,2,3)$, which
determine the mean-field propagation of the hadrons, and by the
transition rates $G^\dagger G\,\delta^4 (\ldots )$ in the collision
term, that describes the scattering and hadron production/absorption
rates.

The scalar and vector mean fields $U_{h}^{S}$ and $U^\mu_{h}$ for
baryons are taken from Ref.~\cite{Ehehalt} and don't have to be
specified here again. The pions and $\eta$'s as Goldstone bosons are
expected not to change their properties in the medium; they will be
treated as bare particles throughout all calculations. Furthermore, the
$K^+$ meson energy changes only very moderately with baryon density
according to Kaplan and Nelson \cite{Kaplan,Nelson,Liko} due to a
partial cancellation of the scalar and vector kaon selfenergies. Thus,
they are also produced and propagated as free particles. The antikaon
and vector meson potentials in the medium, however, have to be
specified more explicitly.

\subsection{$K^-, \rho, \omega, \Phi$  selfenergies}

As in case of antiprotons there are a couple of models for the antikaon
and vector meson selfenergies which differ in the actual magnitude of
the meson potential.  Without going into a detailed discussion of the
various approaches we adopt the more practical point of view, that the
actual $K^-, \rho, \omega$ and $\Phi$ selfenergies are unknown and as a
guide for our analysis use a linear extrapolation of the form (for the
meson $x$),
\begin{equation}
\label{kmass}
m^*_x(\rho_B) = m_x^0 \left(1 - \alpha_x \frac{\rho_B}{\rho_0}\right) \geq m_q + 
m_{\bar{q}},
\end{equation}
with $\alpha_x \approx $ 0.2 for antikaons, which leads to a fairly
reasonable reproduction of the antikaon mass from
Refs.\cite{Kaplan,Nelson} and the recent results from Waas, Kaiser and
Weise \cite{waas}, $\alpha_x \approx 0.18$ for $\rho$ and $\omega$
mesons according to Hatsuda and Lee \cite{hatsuda} and $\alpha_x
\approx 0.025$ for the $\Phi$ meson.  We note that the dropping of the
meson masses is associated with a corresponding scalar energy density
in the baryon/meson Lagrangian, such that the total energy-momentum is
conserved during the heavy-ion collision (cf. Ref.~\cite{Ehehalt}).

\subsection{Elastic and inelastic reaction channels}

Baryon-baryon (BB) collisions are described using free differential
cross sections from Ref.~\cite{landolt} for invariant energies $\sqrt{s}
\leq$ 2.6 GeV and by the LUND string formation and fragmentation model
\cite{lund} for $\sqrt{s} \geq$ 2.6 GeV, which generates the hadronic
final states of a BB collision. The same concept is used for
meson-baryon (mB) reactions, where for $\sqrt{s} \leq$ 2.3 GeV
differential cross sections from Ref.~\cite{landolt} are employed
whereas the LUND model is used for higher invariant energies.
Meson-meson (mm) reactions (e.g. $\pi \pi \rightarrow \rho, \pi \rho
\rightarrow \Phi, \pi \rho \rightarrow a_1$) are described within the
Breit-Wigner resonance picture using branching ratios from the nuclear
data tables \cite{PDB}. In all reaction channels the thresholds are
shifted according to the actual mass of the hadrons (at finite baryon
density $\rho_B = \sqrt{j_\mu j^\mu}$, where $j_\mu(x)$ is the local
baryon current).  Also within the Breit-Wigner resonance formation the
actual masses of the hadrons are used, whereas their width is corrected
according to the local phase space for the decay. For the detailed
parametrizations employed (including also baryon-hyperon and
meson-hyperon channels) the reader is refered to
Refs.~\cite{Ehehalt,Teis96,Cass96}.

\subsection{Optimizing for high baryon density}

\begin{figure}[h1]
\vspace*{-10.5cm}\hspace*{-25mm}
\psfig{figure=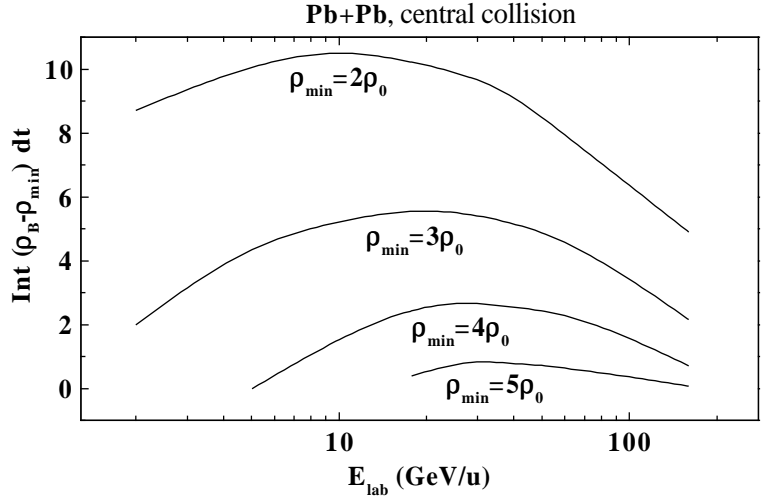,width=200mm,height=300mm}
\vspace*{-13.2cm}
\caption{ The quantity F (\ref{e1}) for central Pb + Pb collisions as a
function of the bombarding energy per nucleon for 4 different cuts in
$\rho_{min}$.}
\label{Fig1}
\end{figure}

In order to probe the restoration of chiral symmetry at high baryon
density in nucleus-nucleus collisions, one has to perform experiments
with heavy nuclei (e.g. Pb + Pb) and optimize the beam energy to
achieve a large volume of high baryon density for a sufficiently long
time. In this respect central collisions of Pb + Pb have been
investigated within the transport approach specified above and the
'stopped' baryon density $\rho_b^s(t)$ - including only baryons with
rapidity $|y| \leq $ 0.7 in the cms - has been computed in a central
volume $V = \pi R^3/\gamma_{CM}$ with $R = $ 4 fm, while $\gamma_{CM}$
is the Lorentz factor in the nucleus-nucleus center-of-mass system.
Since we are interested in high baryon densities above some value
$\rho_{min}$ for long times, we consider the quantity
\be
\label{e1}
F = \int dt \ (\rho_B^s(t) - \rho_{min}) \ \Theta(\rho_B^s(t) - \rho_{min}),
\ee
which should serve as a useful guide in the optimization problem. The
quantity F (\ref{e1}) is displayed in Fig. 1 for central collisions of
Pb + Pb from 1 - 200 GeV/u for different values of $\rho_{min}$ from 2
- 5 $\rho_0 \ (\rho_0 \approx $0.168$  fm^{-3})$. Thus optimal
bombarding energies for baryon densities above 4$ \rho_0$ should be
around 20 - 30 GeV/u in order to explore the properties of an
intermediate phase, where the chiral symmetry might approximately be
restored and the hadron masses (except for the Goldstone bosons) might
be close to their current quark masses $m_q + m_{\bar{q}}$.

\section{Meson production in nucleus-nucleus collisions}

\subsection{Pions}

The pions as the lightest Goldstone bosons are not expected to change
their properties in the dense nuclear medium significantly - except in
a quark-gluon-plasma (QGP) phase - such that their production and
propagation should be reasonably described without introducing any
selfenergies. As an example for SIS energies we show in Fig.~2 the
calculated results \cite{Ehehalt} for transverse $\pi^0$ spectra in Ar
+ Ca collisions at 1.5 GeV/u (for 0.68 $\leq y_{lab} \leq$ 0.84) in
comparison to the data of the TAPS collaboration \cite{Metag} (open
squares), which are described over four orders of magnitude with
relative deviations of less than 30\%. Similar results have been
obtained for various systems at 1 - 2 GeV/u in Ref.~\cite{Teis96}.
\begin{figure}[h1]
\vspace*{-13.5cm}\hspace*{-2cm}
\psfig{figure=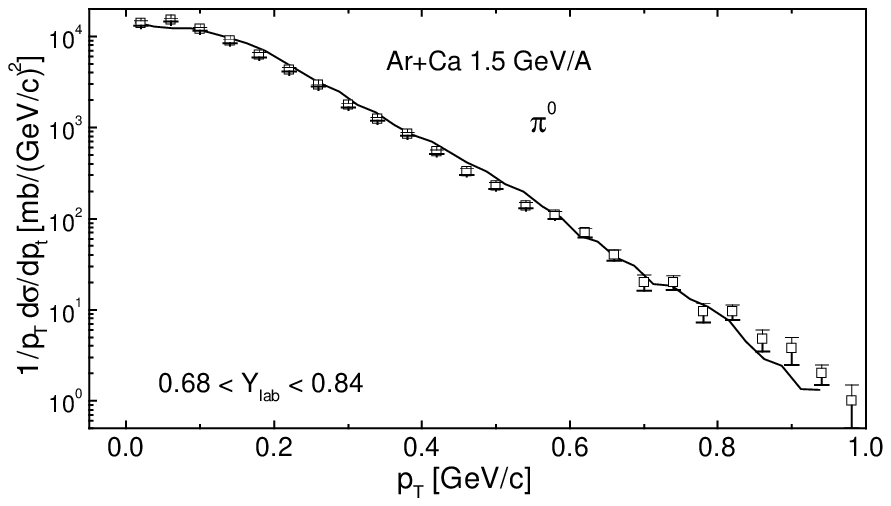,width=220mm,height=340mm}
\vspace*{-15.2cm}
\caption{ The calculated transverse $\pi^0$ spectra (solid line)
in Ar + Ca collisions
at 1.5 GeV/u (for 0.68 $\leq y_{lab} \leq$ 0.84) in comparison to the data of
the TAPS collaboration\cite{Metag} (open squares)}
\label{Fig2}
\vspace*{-11cm}\hspace*{-1.5cm}
\psfig{figure=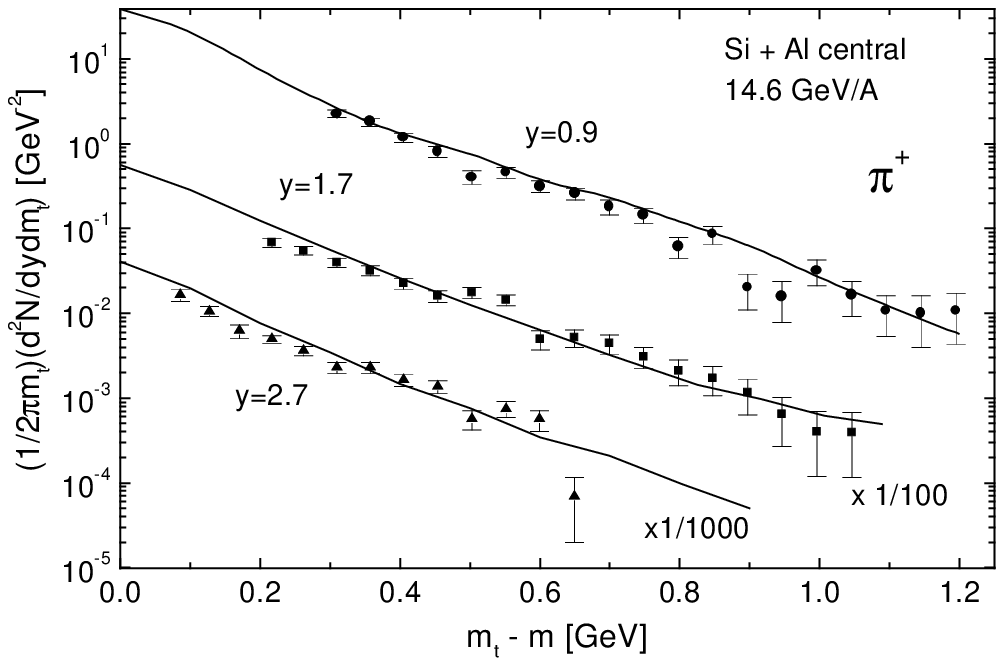,width=200mm,height=280mm}
\vspace*{-11.2cm}
\caption{ The calculated transverse mass spectra for  $\pi^+$ (solid
lines) in ntral collisions of Si + Al at 14.6 GeV/u for  different
rapidities in laboratory $(y = 0.9, 1.7, 2.7)$ in comparison to the
data from Ref.~\protect\cite{abbott}.}
\label{fig3}
\end{figure}
Quantitatively similar experiences have been made at AGS energies as
can be extracted from Fig. 3 where the transverse $\pi^+$ mass spectra
for central collisions of Si + Al at 14.6 GeV/u are displayed for 3
different rapidities in the laboratory system in comparison to the
data  from Ref.~\cite{abbott}.

\noindent
As an example for the pion yield at SPS energies we show in Fig. 4 the
$\pi^-$ rapidity distribution for central S + S collisions at 200 GeV/u
in comparison to the data from Ref.~\cite{ss} (open squares), which are
approximately of Gaussian shape. At midrapidity (y = 0) here the
$\pi^-$ density is about a factor of 7 higher than the proton density.
This indicates that the available energy is dominantly used for mass
production (in form of pions) and that during the longitudinal
expansion of the 'hadronic fireball' meson-meson reaction channels
should occur more frequent than meson-baryon or baryon-baryon
reactions.

\begin{figure}[h1]
\vspace*{-13.5cm}\hspace*{-2cm}
\psfig{figure=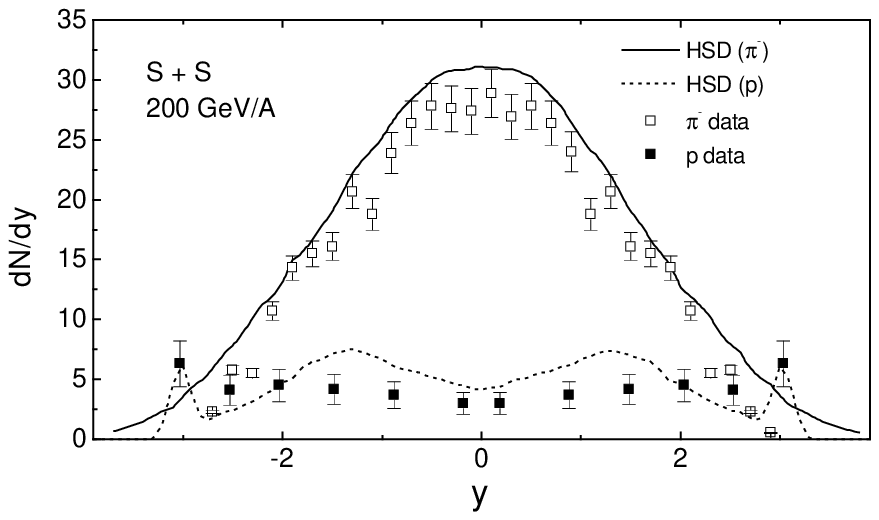,width=220mm,height=340mm}
\vspace*{-15.2cm}
\caption{The $\pi^-$ rapidity distribution (solid line) for central S +
S collisions at 200 GeV/u in comparison to the data from
Ref.~\protect\cite{ss} (open squares).  The dotted line shows the
calculated proton rapidity distribution in comparison to the respective
data from Ref.~\protect\cite{ss} (full squares).}
\label{Fig4}
\vspace*{-13.2cm}\hspace*{-2cm}
\psfig{figure=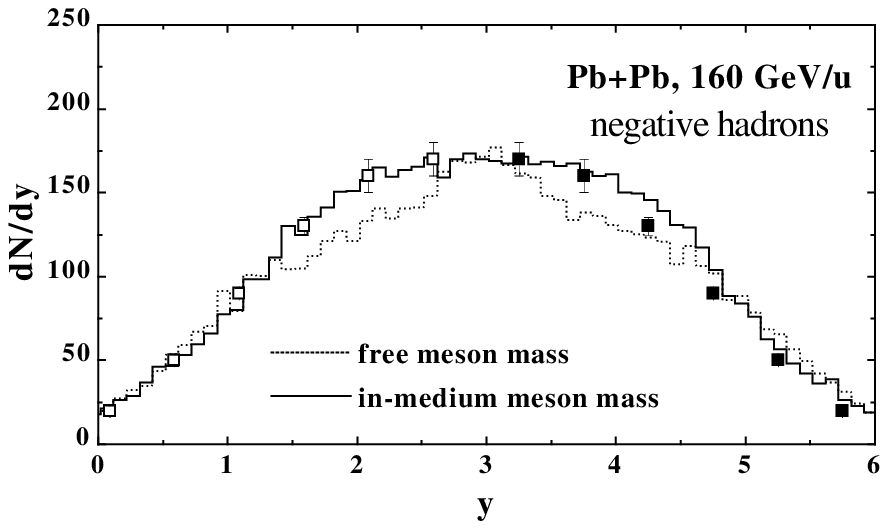,width=22cm,height=340mm}
\vspace*{-15.3cm}
\caption{The preliminary rapidity distribution of negative hadrons from
NA49 \protect\cite{NA49} (full squares) in comparison to the HSD
results \protect\cite{brat96}. The solid line corresponds to a
calculation (at b = 2fm) including the dropping meson masses from
Eq.~(\protect\ref{kmass}), whereas the dotted line results from a
calculation with bare meson masses.  The open squares are obtained by
reflecting the full squares at midrapidity.}
\label{Fig5}
\end{figure}

The shape of the pion rapidity distribution is not changed
significantly when going over to central collisions of Pb + Pb at 160
GeV/u as can seen from Fig. 5, where the preliminary rapidity
distribution of negative hadrons (essentially $\pi^-$, $K^-$ and
$\bar{p}$) from NA49 \cite{NA49} is shown in comparison to the HSD
results \cite{brat96}.  Here the solid line corresponds to a
calculation (at b = 2fm) including the dropping meson masses from
Eq.~(\ref{kmass}), whereas the dotted line results from a calculation with
bare meson masses. The broadening of the rapidity distribution around
midrapidity (y $\approx$ 3) in the dropping mass scenario is due to
pions from $\rho$ and $\omega$ decays, which are produced with a wider
distribution in rapidity in this case. Since the proton distribution
$dN_p/dy \approx 40 - 45$ at midrapidity (cf. Fig. 22 in
Ref.~\cite{Ehehalt}) the $\pi^-$ to proton ratio is only about 4 for
central Pb + Pb collisions such that the baryon density in the
expanding 'hadronic fireball' is significantly higher than that for
central S + S collisions.

\subsection{Kaons and antikaons}

Whereas kaons should feel a slightly repulsive potential in the nuclear
medium according to the approach by Kaplan and Nelson \cite{Kaplan} or
Waas, Kaiser and Weise \cite{waas}, the antikaons should experience a
stronger attractive potential at finite baryon density, which is also
supported by $K^-$ atomic data \cite{kmin}.  As a first order
approximation we thus assume the $K^+$ potential or selfenergie to be
zero and adopt the linear parametrization for the in-medium $K^-$
mass from Eq.~(\ref{kmass}).  For the detailed reaction channels and
cross sections considered the reader is refered to Ref.~\cite{Cass96}.

The l.h.s. of Fig. 6 shows the calculated results for the inclusive
$K^+$ invariant cross section for Ni~+~Ni collisions at 0.8, 1.0 and
1.8 GeV/u at $\theta_{lab} = $44$^o$, that have been transformed to the
nucleus-nucleus cms, in comparison to the preliminary $K^+$ spectra from
Ref.~\cite{Senger,Psenger}. Since the data can be described quite
reasonably at all energies from 0.8 - 1.8 GeV/u, apparently no
selfenergy effects are needed for $K^+$ mesons.  This finding is also
in accordance with earlier studies on $K^+$ production in
nucleus-nucleus~\cite{lang,Tomo,Cass90} and proton-nucleus
collisions~\cite{ca90a}. On the hand, due to the rather stable
quasi-particle properties of the kaons in the medium, they qualify as
probes in connection with the nuclear-equation-of-state (EOS) as
suggested early by Aichelin and Ko \cite{Aichelin2}.

The r.h.s. of Fig. 6 shows the calculated $K^-$ spectra for Ni~+~Ni at
1.85 GeV/u at $0^o$ with respect to the beam axis in the
nucleus-nucleus cms in comparison with the data of Ref.~\cite{Schro}
(full squares) and the preliminary data for Ni~+~Ni at 1.8 GeV/u from
Ref.~\cite{Senger1} (open dots).  The dashed line reflects a
calculation including the bare antikaon mass without any antikaon
absorption, while the dash-dotted line includes antikaon absorption,
which reduces the cross section on average by a factor of 5 for the
Ni~+~Ni system.  However, the data are underestimated severely in the
bare $K^-$ mass approximation. The solid line in Fig. 6 (r.h.s.) shows
the result of a calculation, where the $K^-$ mass drops with baryon
density according to Eq.~(\ref{kmass}) with $\alpha_x$ = 0.2 including
also antikaon reabsorption\footnote{For practical purposes one should
consider $\alpha_x $ to be a free parameter to be fixed in comparison
to the experimental data in order to learn about the magnitude of the
antikaon selfenergy.  In fact, we obtain a much better reproduction of
the spectra for $\alpha_x \approx 0.24$, but due to the uncertainties
involved in the elementary $BB$ production cross sections one cannot
determine this value very reliably.}.  With increasing $\alpha_x$ not
only the magnitude of the spectrum is increased, but also the slope
becomes softer. For $\alpha_x \approx 0.2$ we still underestimate the
experimental spectra slightly, but it is clearly seen that quite
sizeable antikaon attractive selfenergies are needed to reproduce the
data. This finding is also in line with an independent calculation by
Li, Ko and Fang \cite{Fang}.

\begin{figure}[h1]
\vspace*{-10cm} \hspace*{-28mm}
\psfig{figure=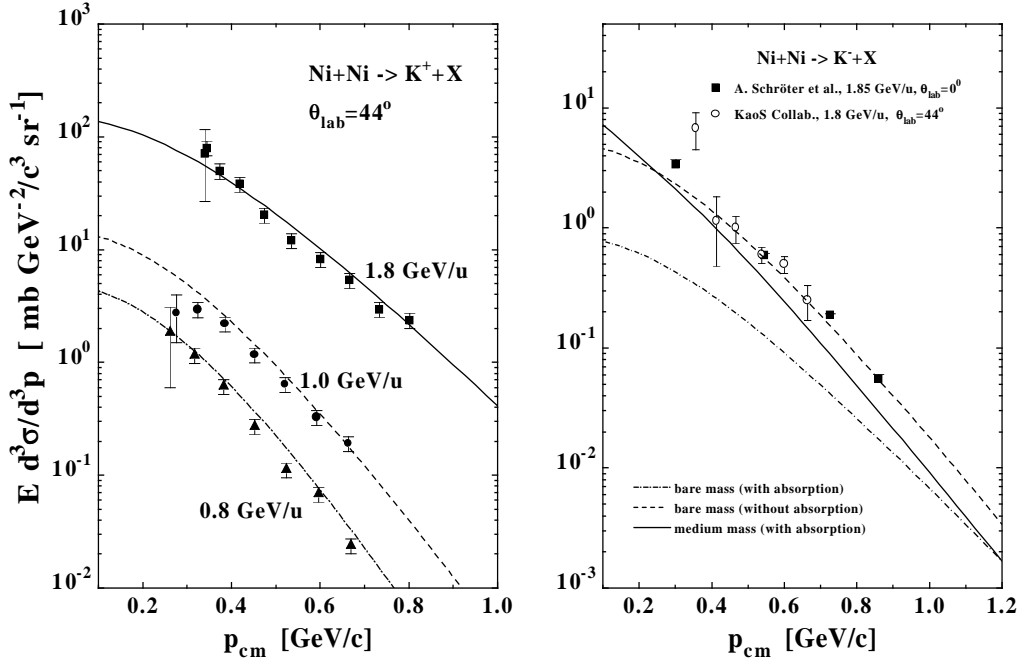,width=200mm,height=300mm}
\vspace*{-11.8cm}
\caption{(l.h.s.) The calculated results for the inclusive $K^+$
invariant cross section for Ni~+~Ni collisions at 0.8, 1.0 and 1.8
GeV/u at $\theta_{lab} = $44$^o$, that have been transformed to the
nucleus-nucleus cms, in comparison to the preliminary $K^+$ spectra from
Ref.~\protect\cite{Senger,Psenger}. (r.h.s.) The calculated $K^-$
spectra for Ni~+~Ni at 1.85 GeV/u at $0^o$ with respect to the beam
axis in the nucleus-nucleus cms in comparison with the data of
Ref.~\protect\cite{Schro} (full squares) and the preliminary data for
Ni~+~Ni at 1.8 GeV/u from Ref.~\protect\cite{Senger1} (open dots).  The
dashed line reflects a calculation including the bare $K^-$ mass
without any antikaon absorption, while the dash-dotted line includes
antikaon absorption. The solid line corresponds to a calculation with a
dropping antikaon mass according to Eq.~(\protect\ref{kmass}) for
$\alpha_x$ = 0.2.}
\label{Fig6}
\end{figure}

The enhanced production of strangeness is also known from experiments
at AGS and SPS energies\cite{review}. As an example Fig. 7 shows the
measured $K^+/\pi^+$ ratio for pp, Si + Al, Si + Cu, Si + Au and Au +
Au at AGS energies \cite{kags}, which increases by about a factor of 3
with the number of participating nucleons. Whereas a HSD calculation
with in-medium meson masses (solid line) \cite{Ehehalt} approximately
reproduces this trend, the same calculation with bare meson masses
(lower dotted line) underestimates the $K^+/\pi^+$ ratio significantly.
The actual enhancement in the dropping mass scenario is due to a large
contribution from the channel $meson + meson \rightarrow K^+ K^-$,
which is enhanced sizeably for a dropping antikaon mass. Differential
transverse momentum spectra for kaons and antikaons at midrapidity for
central Au + Au collisions at 10.8 GeV/u should shed some further light
on this issue. Note, that RQMD calculations \cite{agsrqmd} for these
systems also underestimate the $K^+/\pi^+$ ratio by about a factor of 1.5.

\begin{figure}[h1]
\vspace*{-13.2cm}\hspace*{-2.5cm}
\psfig{figure=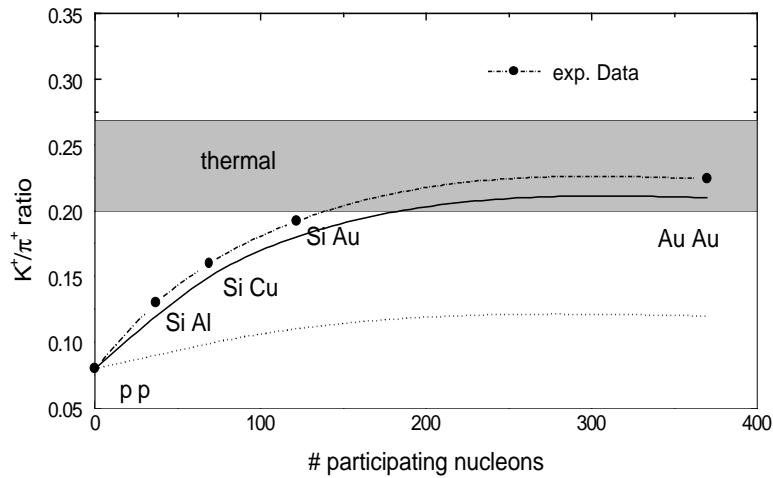,width=220mm,height=350mm}
\vspace*{-15.5cm}
\caption{ The measured $K^+/\pi^+$ ratio for pp, Si + Al, Si + Cu, Si +
Au and Au + Au at AGS energies \cite{kags} (full dots) in comparison to
HSD calculations with in-medium meson masses (solid line)
\cite{Ehehalt} and bare meson masses (lower dotted line).}
\label{Fig7}
\end{figure}

\section{Electromagnetic probes}

Photons and dileptons are particularly well suited for an investigation
of the violent phases of a high-energy heavy-ion collision because they
can leave the reaction volume essentially undistorted by final-state
interactions. Whereas the signal from direct photons is largely covered
by the electromagnetic decays of light neutral mesons ($\pi^0, \eta$),
$e^+e^-$ or $\mu^+ \mu^-$ pairs at higher invariant masses do not
suffer that much from large background contributions. Indeed, dileptons from
heavy-ion collisions  have been observed by the DLS collaboration at
the BEVALAC \cite{ro88,na89,ro89} and by the CERES \cite{CERES}, HELIOS
\cite{HELIOS,HELI2}, NA38 \cite{NA38} and NA50 \cite{NA50}
collaborations at SPS energies.

\subsection{$e^+e^-$ pairs}

Quite some years ago it has been found within microscopic transport
studies at BEVALAC/SIS energies \cite{Wolf} that above about 0.5 GeV of
invariant mass (of the lepton pair) the dominant production channel is from
$\pi^+ \pi^-$ annihilation, such that the properties of the short
lived $\rho$ meson could be explored at high baryon density. The data
available so far, however, did not allow for a closer distinction of
the various models proposed.  At SPS energies the enhancement of the
low mass dimuon yield in S + W compared to p + W collisions
\cite{HELIOS} has been first suggested by Koch {\it et al.}\cite{Koch}
to be due to $\pi^+\pi^-$ annihilation.  Furthermore, Li {\it et
al.}\cite{LKB} have proposed that the enhancement of the $e^+e^-$ yield
in S + Au collisions - as observed by the CERES collaboration
\cite{CERES} - should be due to an enhanced $\rho$-meson production
(via $\pi^+\pi^-$ annihilation) and a dropping $\rho$-mass in the
medium. In fact, their analysis - which was based on an expanding
fireball scenario in chemical equilibrium - could be confirmed within
the HSD transport calculations in Ref.~\cite{Cass95}.  However, also a
more conventional approach including the increase of the $\rho$-meson
width in the medium due to the coupling of the $\rho, \pi, \Delta$ and
nucleon dynamics \cite{Herrmann} was found to be roughly compatible
with the CERES data. On the other hand, the dimuon data of the HELIOS-3
collaboration\cite{HELIOS} could only be described satisfactorily when
including dropping meson masses\cite{Ca96}. In the following some more
recent results from Refs.~\cite{brat96a,brat96} are reported, where
systematic studies on the various dilepton channels from 10 - 200 GeV/u
have been performed.

\begin{figure}[h1]
\vspace*{-17.8cm} \hspace*{-5.cm}
\psfig{figure=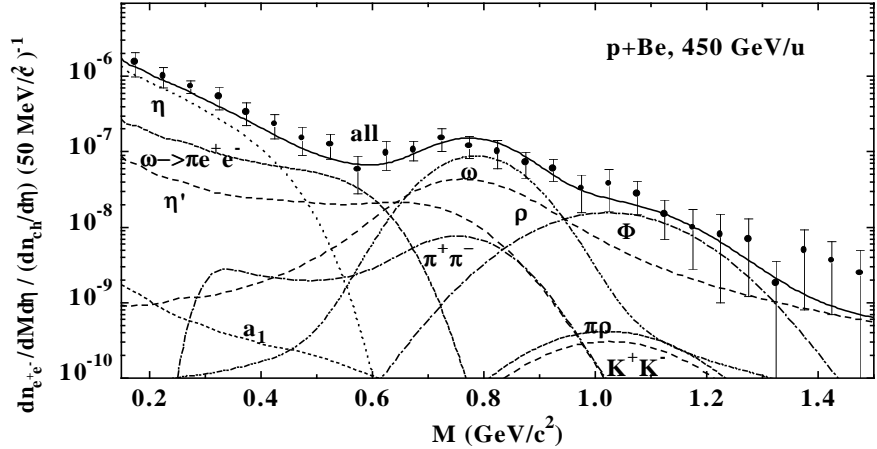,width=250mm,height=350mm}
\vspace*{-12.1cm}
\caption{  The calculated dilepton spectra (full solid line)
for p + Be at 450 GeV/u in comparison with the data from Ref.~\cite{CERES}.
The thin lines indicate the individual contributions from the different
production channels including the CERES-acceptance and mass resolution.}
\label{Fig8}
\end{figure}

As an example for dilepton spectra at SPS energies Fig. 8 shows the
spectral decomposition as a function of the $e^+e^-$ invariant mass M
for p + Be at 450 GeV/c in comparison to the data of the CERES
collaboration \cite{CERES}. In this case the spectrum can be fully
accounted for by the electromagnetic decays of the $\eta, \eta'$ and
vector mesons $ \rho^0, \omega$ and $\Phi$. Contributions from
meson-meson channels ($\pi^+\pi^-, K^+K^-, \pi \rho$) are of minor
importance.
\begin{figure}[t]
\vspace*{-5cm}\phantom{a}\hspace{-9cm}
\psfig{figure=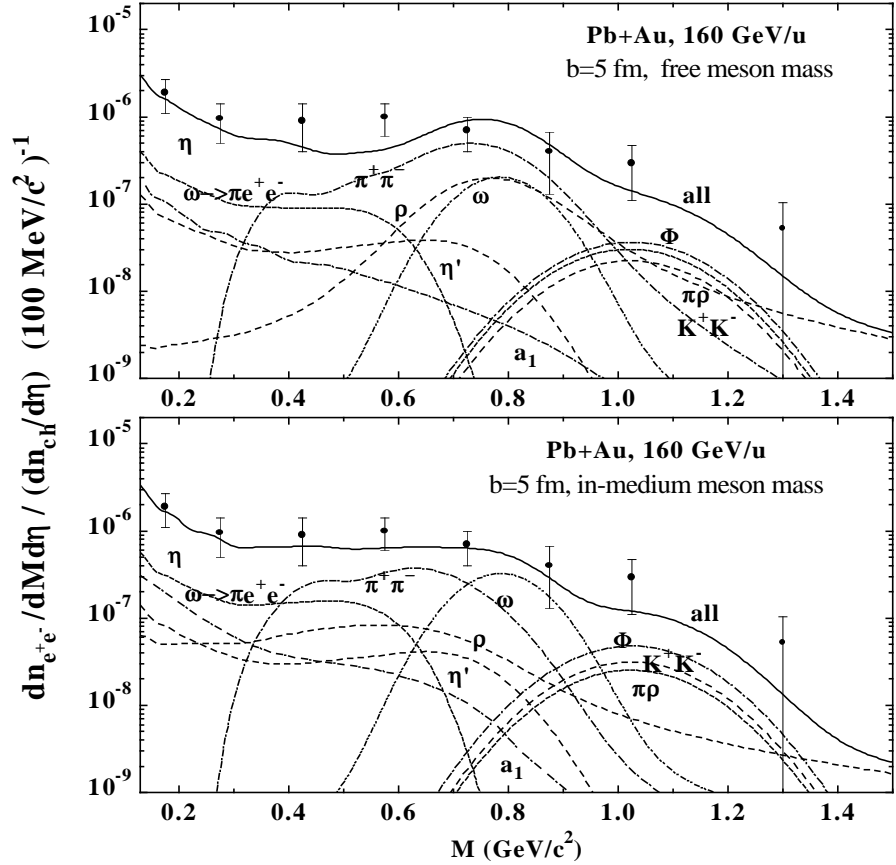,width=250mm,height=350mm}
\vspace*{-19cm}
\caption{Dilepton invariant mass spectra for semicentral collisions of
Pb + Au at 160 GeV/u (full solid lines) in comparison to the
preliminary data of the CERES collaboration \protect\cite{Ulrich}. The
upper part shows the results of a calculation with bare meson masses
whereas the lower part includes the dropping meson masses
(\protect\ref{kmass}).}
\label{Fig9}
\end{figure}

The situation changes quite dramatically when going over to
nucleus-nucleus collisions. For Pb + Au at 160 GeV/u (and semicentral
collisions) the dominant yield for invariant masses 0.3 GeV $\leq$ M
$\leq$ 0.7 GeV stems from $\pi^+\pi^-$ annihilation (cf. Fig. 9). Also
in the $\Phi$ mass regime about 1 GeV there is a large contribution from
$K^+K^-$ and $\pi \rho$ annihilation to dileptons for both
scenarios: with bare meson masses (upper part of Fig. 9) and with
in-medium meson masses (lower part of Fig. 9).  Whereas most of the
processes (Dalitz and direct decays) occur in the vacuum at zero baryon
density, the $\pi \pi \rightarrow \rho^0 \rightarrow e^+e^-$ and direct
$\rho^0$ (from BB and mB collisions) decay still occur at finite baryon
density such that a dropping $\rho$ mass also leads to a shift of the
respective contribution to lower invariant masses M. In Fig. 9 both
scenarios are compared to the preliminary data of the CERES
collaboration \cite{Ulrich} including the experimental cuts in rapidity
y, transverse momentum of the leptons as well as the CERES mass
resolution; due to the present statistics, however, there is no unique
conclusion since the calculation with bare meson masses (upper part)
also describes the data except for one point at 0.6 GeV. On the other
hand, the present preliminary data match well with the calculation
including the in-medium meson masses. Apart from better statistics also
a higher mass resolution of the CERES detector (especially in the
$\rho, \omega, \Phi$ mass regime) should allow to disentangle the
different scenarios in the next years.

\section{Summary}

In this article the production of secondary particles in proton-nucleus
and nucleus-nucleus collisions from 1 - 200 GeV/u has been investigated
within the covariant transport approach HSD \cite{Ehehalt}.  The
analysis shows that $\pi$ and $K^+$ spectra are reasonably  well
described in this energy regime without introducing any medium
modifications for these mesons (cf. also Ref.~\cite{Teis96} in case of
pions).  This experience is fully in line with earlier studies on this
subject and the results from independent groups~\cite{Hartnack,Bass}.
The antikaon spectra, however, are underestimated severely when
incorporating only bare kaon masses roughly in line with the study by
Li {\it et al.}\cite{Fang}.  When including an attractive antikaon
potential comparable to that proposed by Waas, Kaiser and Weise
\cite{waas}, a satisfactory description of the $K^-$ spectra can be
given, both in the actual magnitude as well as in the slope. It is
worth noting that the $\pi$-hyperon $\rightarrow K^- N$ production
channels play a sizeable role in case of the vacuum antikaon mass,
whereas their contribution  in the 'dropping mass scenario' becomes of
minor importance.

The 'observed' dropping of the antikaon mass with baryon density may be
interpreted as a step towards a partial restoration of chiral symmetry
that can already be seen at SIS energies (1 - 2 GeV/u).  Similar
observations have been made at AGS energies (10 - 15 GeV/u) (cf. Fig.
7) as well as at SPS energies (160 - 450 GeV/u), where especially a
dropping of the $\rho$-mass can be used to accurately describe the
dilepton spectra from heavy-ion reactions~\cite{Cass95,LKB,Ca96} at
invariant masses 0.3 GeV $\leq$ M $\leq$ 0.7 GeV. The enhancement of
dileptons (about a factor of 3) seen in the $\Phi$ mass region is
essentially due to the meson-meson production channels $\pi \rho
\rightarrow \Phi$ and $K^+K^- \rightarrow \Phi$.

Though there are quite a number of indications for dropping meson
masses in the medium by now, one has to properly examine the
possibility that conventional many-body effects such as
'resonance-hole' loops~\cite{friman} may also account for the spectra
observed.  In addition, detailed experimental studies on $K^-, \rho$
and $\omega$ production in proton (pion) - nucleus reactions should be
performed close to threshold energies since a 20\% reduction of their
mass at density $\rho_0$ should clearly be visible in the respective
spectra \cite{Leonid}.

\section*{Acknowledgments}
The author acknowledges valuable and inspiring discussions with A.~Drees,
V.~Metag, P.~Senger, H.~Specht and Gy.~Wolf. Furthermore, he
likes to thank his collegues and collaborators for their help and work,
on which many of the results presented in this article are based, in
particular E.~L.~Bratkovskaya, W.~Ehehalt, C.~M.~Ko, U.~Mosel,
A.~Sibirtsev and S.~Teis.

\end{document}